\documentclass[pra,preprint,a4paper]{revtex4}
\usepackage{graphicx}
\usepackage{palatino}
\usepackage[latin1]{inputenc}
\usepackage{bbm}
\usepackage{amsmath,amssymb}
\usepackage{latexsym}
\usepackage{calc}
\usepackage{shadow}

\newcommand{\be}{\begin{equation}}
\newcommand{\beq}{\begin{equation}}
\newcommand{\ee}{\end{equation}}
\newcommand{\bea}{\begin{eqnarray}}
\newcommand{\eea}{\end{eqnarray}}
\newcommand{\ba}{\begin{array}}
\newcommand{\ea}{\end{array}}

\renewcommand{\vr} {{\bf r}}
\newcommand{\vj} {{\bf j}}
\newcommand{\vs} {{\bf s}}

\newcommand{\nn} {\nonumber}

\begin{document}
\title{Orbital currents in the Colle-Salvetti correlation energy functional and the degeneracy problem}
\author{S. Pittalis$^1$, S. Kurth$^1$, S. Sharma$^{1,2}$}
\author{E.K.U. Gross$^1$}
\affiliation{1 Institut f\"{u}r Theoretische Physik, Freie Universit\"at Berlin,
Arnimallee 14, D-14195 Berlin, Germany}
\affiliation{2  Fritz Haber Institute of the Max Planck Society, Faradayweg 4-6, 
D-14195 Berlin, Germany.}

\date{\today}

\begin{abstract}
Popular density functionals for the exchange-correlation energy typically
fail to reproduce the degeneracy of different ground states of open-shell atoms.
As a remedy, functionals which explicitly depend on the current density have
been suggested. We present an analysis of this problem by investigating
functionals that explicitly depend on the Kohn-Sham orbitals. Going beyond
the exact-exchange approximation by adding correlation in the form of the
Colle-Salvetti functional we show how current-dependent terms enter the
Colle-Salvetti expression and their relevance is evaluated.
A very good description of the degeneracy of ground-states for atoms of
the first and second row of the periodic table is obtained.
%In
%spin-unrestricted calculations absolute errors less than 1 kcal/mol, with  
%average of 0.5 kcal/mol, are observed.
\end{abstract}

\maketitle

\section{Introduction}

Common approximations to the exchange-correlation functional of
density functional theory (DFT) \cite{HohenbergKohn:64,KohnSham:65}
and spin-DFT (SDFT) \cite{BarthHedin:72}
often fail to reproduce the degeneracy of different ground states.
An illustrative  example are ground states of open-shell atoms where one usually and
erroneously obtains different total energies for states with
zero and non-vanishing current density.
The local spin density approximation (LSDA) 
gives rather small splittings (of the order of 
1 kcal/mole), but generalized gradient approximations (GGAs) and 
meta-GGAs can introduce splittings of 10 kcal/mol 
\cite{ZieglerRaukBaerends:77,Barth:79,KutzlerPainter:91,MerkleSavinPreuss:92,Baerends:97,Becke:02,MaximoffErnzerhofScuseria:04,TaoPerdew:05}.

These spurious energy splittings would vanish 
if the exact exchange-correlation functional could be used.
The exchange-correlation energy functional can be represented in terms
of the exchange-correlation hole function.
Considering current-carrying states, Dobson showed how the expression for the
exchange-hole curvature 
has to be changed by including current-dependent terms \cite{Dobson:93}. Later, Becke 
found the same kind 
of terms in the short-range behavior of the 
exchange-correlation hole, and observed that
they also enter the spin-like correlation-hole
function \cite{BeckeJ:96}. For open-shell atoms, inclusion of these current-dependent terms
results in spurious energy splittings
of less than 1 kcal/mole \cite{Becke:02}.
Along these lines, Maximoff at al. \cite{MaximoffErnzerhofScuseria:04} worked 
out a correction for the system-averaged exchange hole of the 
Perdew-Burke-Ernzerhof (PBE) GGA
\cite{PerdewBurkeErnzerhof:96}, which improves
the corresponding spurious splittings.
Alternatively, Tao and Perdew \cite{Tao:05, TaoPerdew:05} proposed a scheme for the extension of 
existing functionals using ideas of current density functional theory (CDFT)
\cite{VignaleRasolt:87,VignaleRasolt:88} which, again, improves the description of the 
degeneracy. 

The performance of the 
exact-exchange (EXX) energy functional - which, by definition, describes 
the exchange hole correctly - has been evaluated  for the spurious splittings
in DFT and SDFT \cite{PittalisKurthGross:06}. 
In the EXX-DFT (i.e., spin-restricted calculations using one and the same Kohn-Sham potential for spin-up and spin-down orbitals)
the degeneracy is well reproduced to 
within 0.6 kcal/mole but, surprisingly, in EXX-SDFT (i.e., spin-unrestricted calculations using two Kohn-Sham potentials, one for spin-up and
one for spin-down orbitals)
spurious splittings up to 3 kcal/mole are obtained. In particular, current-carrying states
always have higher total energies than states without current.  
This observation motivated the applications of the optimized-effective-potential (OEP) method
\cite{SharpHorton:53,TalmanShadwick:76,GraboKreibichKurthGross:00} generalized to 
current-spin-density functional theory (CSDFT) 
to these current-carrying states \cite{PittalisKurthHelbigGross:06}. 
As expected, EXX-CSDFT total energies for current-carrying states are lower than those of EXX-SDFT. 
However, this lowering 
is too small to give a substantial improvement of the spurious energy splittings. 
These studies lead to the conclusion that correlation is needed for 
any further improvement.

The construction of a correlation energy functional compatible with EXX
is a difficult task \cite{IvanovHirataBartlett:99,Goerling:99,SeidlPerdewKurth:00-2}, 
but for spherical atoms it was found that EXX
combined with the Colle-Salvetti (CS) functional for correlation 
\cite{ColleSalvetti:75,ColleSalvetti:79,ColleSalvetti:83} leads to very 
accurate total energies \cite{GraboGross:95}. 
The CS functional has been used to derive the popular Lee-Yang-Parr (LYP) functional \cite{LeeYangParr:88}, which is 
most commonly used together with Becke's exchange functional \cite{Becke:88} 
(BLYP) and in hybrid schemes such as B3LYP \cite{Becke:93,Becke:93-2}. 
On the other hand, the CS correlation energy functional also has its 
limitations \cite{SinghMassaSahni:99,TaoGoriPerdewMcWeeny:01,ImamuraScuseria:02}. 
In particular, while short-range correlations are well described 
\cite{TaoGoriPerdewMcWeeny:01}, very important long-range correlations are 
missing. These correlations often cannot be ignored in molecules and solids, 
but are negligible in atoms. This fact, together with
the encouraging results for spherical atoms \cite{GraboGross:95}, indicates that it is appropriate 
to employ the CS functional to analyze the degeneracy problem for open-shell atoms 
beyond EXX. Furthermore, the expression of the CS functional also allows a 
reconsideration of the relevance of the orbital currents 
as ingredient of correlation functionals.

Although the general density functional 
formalism to deal with degenerate 
ground states includes densities which can only be obtained by a weighted sum 
of several determinantal densities \cite{UllrichKohn:01,UllrichKohn:02}, 
as in many previous investigations \cite{Baerends:97,Becke:02,MaximoffErnzerhofScuseria:04,TaoPerdew:05}
we only consider densities which 
may be represented by a single Slater determinant of Kohn-Sham orbitals.

\section{Theory}

Going beyond the EXX approximation, we here consider the correlation-energy functional
of Colle and Salvetti \cite{ColleSalvetti:75}. This expression relies on the 
assumption that the correlated two-body reduced density matrix may be 
approximated by the Hartree-Fock (HF) two-body reduced density matrix
$\rho_2^{HF}(\vr_1,\vr_2)$, multiplied by a Jastrow-type correlation factor. After a series of 
approximations, the following expression is obtained for the correlation 
energy 
\be\label{ocs}
E_c=-4a 
\int d\vr \; 
\frac{ \rho_2^{HF}(\vr,\vr) }{ \rho(\vr) } 
\left\{ 
\frac{ 1+b\rho^{-\frac{8}{3}}(\vr)
\left[ \nabla_s^2 \rho_2^{HF}\left(\vr+\frac{\vs}{2}\; ,\vr-\frac{\vs}{2} \;
\right) \bigg\vert_{s=0} \right]
e^{ -c\rho^{-\frac{1}{3}}(\vr) } }{1+d\rho^{-\frac{1}{3}}(\vr)}  
\right\}
\ee
where $\rho_2^{HF}(\vr,\vs)$ is expressed 
in terms of the average and relative coordinates $\vr=\frac{1}{2}(\vr_1+\vr_2)$ and 
$\vs=\vr_1-\vr_2$. Here, $\rho(\vr)$ is the electron density and 
the constants $a=0.049$, $b=0.132$, $c=0.2533$, $d=0.349$ are determined by 
a fitting procedure using the Hartree-Fock (HF) orbitals for the Helium atom. 

Following Lee, Yang and Parr, this expression can be restated as a formula 
involving only the total charge-density, the charge-density of each 
Hartree-Fock orbital and their gradient and Laplacian \cite{LeeYangParr:88}. 
In this derivation, the single-particle orbitals are 
tacitly assumed to be real. We denote the resulting expression as CSLYP. 
In the following, we relax this restriction and 
consider complex orbitals. We then proceed in analogy to the
inclusion of current-dependent terms in the Fermi-hole curvature \cite{Dobson:93,Becke:02}
and in the extension of the electron-localization-function (ELF) \cite{BeckeEdgecombe:90} for 
time-dependent states \cite{BurnusMarquesGross:05}. 
As a consequence, in addition to the term already present in CSLYP expression,
the current densities of the single-particle orbitals appear in the final formula.
In order to obtain this expression, which
in the following will be denoted as JCSLYP, we rewrite the  
Laplacian of the Hartree-Fock (HF) two-body reduced density matrix in 
Eq.(\ref{ocs}) in terms of the original particle coordinates
\be
\nabla_s^2 \rho_2^{HF}\left(\vr+\frac{\vs}{2}\; ,\vr-\frac{\vs}{2} \;
\right) \bigg\vert_{s=0} = \left( \frac{1}{4}\nabla_1^2 + 
\frac{1}{4}\nabla_2^2 - \frac{1}{2} \nabla_1\cdot\nabla_2 \right) 
\rho_2^{HF}(\vr_1,\vr_2) |_{\vr_1=\vr_2} \; .
\ee
where
\be
\rho_2^{HF}(\vr_1,\vr_2)
=
\frac{1}{2} \rho(\vr_1)\rho(\vr_2) - \frac{1}{2} \sum_{\sigma} 
\rho_{1,\sigma}^{HF}(\vr_1,\vr_2)\rho_{1,\sigma}^{HF,*}(\vr_1,\vr_2) \; .
\ee
Here, 
\be
\rho_{1,\sigma}^{HF}(\vr_1,\vr_2) = \sum_{k=1}^{N_\sigma} 
\psi_{k,\sigma}(\vr_1) \psi^*_{k,\sigma}(\vr_2),
\ee
is the first-order HF density matrix (for a single Slater determinant) 
expressed in terms of the single-particle orbitals $\psi_{k,\sigma}(\vr)$. 
The corresponding spin-density is simply given by 
\be
\rho_\sigma(\vr) = \rho_{1,\sigma}^{HF}(\vr,\vr).
\ee
Allowing the single-particle orbitals $\psi_{k,\sigma}(\vr)$ to be complex, a 
given orbital not only gives the contribution 
$\rho_{k,\sigma}(\vr) = | \psi_{k,\sigma}(\vr)|^2$ to the density, but also the 
contribution $\vj_{p~k,\sigma}(\vr)= {\rm Im}\left( \psi_{k,\sigma}(\vr) \nabla 
\psi_{k,\sigma}^*(\vr) \right)$ to the paramagnetic current density which 
is given by
\be
\vj_{p,\sigma}(\vr) = \sum_{k=1}^{N_\sigma} \vj_{p~k,\sigma}(\vr)\; .
\ee
After some straightforward algebra, the Laplacian of the second-order HF 
reduced density matrix takes the final form
\bea\label{JCSLYP}
\nabla_s^2 \rho_2^{HF} 
\left(\vr+\frac{\vs}{2}\; ,\vr-\frac{\vs}{2} \;
\right)\bigg\vert_{s=0} &=&
\frac{1}{4} \, \rho(\vr)  \nabla^2 \rho(\vr) 
-\frac{1}{4} \left( \nabla \rho(\vr) \right)^2 
-\frac{1}{4} \sum_{\sigma} \rho_\sigma(\vr)  \nabla^2 \rho_\sigma(\vr)  \\ \nn
&& + \frac{1}{4} \sum_{\sigma} \rho_\sigma(\vr) 
\left[ 
\sum_{k=1}^{N_\sigma} \frac{\left( \nabla \rho_{k,\sigma}(\vr) \right)^2}{\rho_{k,\sigma}(\vr)}  \right]  + J(\vr)
\eea
where 
\be\label{J}
J(\vr) =  \sum_{\sigma} \rho_\sigma(\vr) 
\left[ 
- \frac{{\vj_{p\sigma}^2(\vr)}}{\rho_{\sigma}(\vr)} 
+ \sum_{k=1}^{N_\sigma}  \frac{{\vj_{p~k,\sigma}^2(\vr)}}{\rho_{k,\sigma}(\vr)}
\right]
\ee
contains all the current-dependent terms.
Alternatively, Eq.~(\ref{JCSLYP}) may also be 
expressed in terms of the non-interacting kinetic energy density
\be
\tau_{\sigma}(\vr) = \frac{1}{2} \sum_{k=1}^{N_\sigma} | \nabla 
\psi_{k,\sigma}(\vr)|^2  = \frac{1}{8}
\sum_{k=1}^{N_\sigma} \frac{\left( \nabla \rho_{k,\sigma}(\vr) \right)^2}{\rho_{k,\sigma}(\vr)} +
\frac{1}{2} \sum_{k=1}^{N_\sigma}  \frac{{\vj_{p~k,\sigma}^2(\vr)}}{\rho_{k,\sigma}(\vr)}
\ee
as 
\bea \label{JCSLYP2}
\nabla_s^2 \rho_2^{HF}
\left(\vr+\frac{\vs}{2}\; ,\vr-\frac{\vs}{2} \;
\right) \bigg\vert_{s=0} &=& 
\frac{1}{4} \, \rho(\vr)  \nabla^2 \rho(\vr) 
-\frac{1}{4} \left( \nabla \rho(\vr) \right)^2 
-\frac{1}{4} \sum_{\sigma} \rho_\sigma(\vr)  \nabla^2 \rho_\sigma(\vr)  \nn \\
&& + \sum_{\sigma} 
\left( 2 \rho_\sigma(\vr) \tau_{\sigma}(\vr) - \vj_{p \sigma}(\vr)^2 \right).
\eea

Comparison of Eq.(\ref{JCSLYP}) and Eq.(\ref{JCSLYP2}) shows that
$J(\vr)$, as defined in Eq.(\ref{J}), also contains current-dependent terms coming 
from the kinetic energy density. Thus, this would also suggest to reconsider the gradient 
expansion of $\tau$ for current-carrying states. 
Important consequences maybe expected for all approximated exchange-correlation 
functional involving the kinetic energy density as ingredient, such as the CS 
functional, and meta-GGAs. This issue will be specifically considered in a 
future work. In the next section, we assess the performance of the CS functional, 
and, in particular, the relevance of $J(\vr)$, in reproducing the degeneracy of 
atomic states.

\section{Results and discussion}

We consider ground states of open-shell atoms
having densities that can be represented by a
single Slater determinant of Kohn-Sham orbitals.
Due to the symmetry of the 
problem, these Slater determinants are eigenstates of the $z$-component 
of both spin and orbital angular momentum. The single-particle orbitals
are generally complex-valued and states with different 
total magnetic quantum numbers, $M_L$, correspond to different 
current densities. By means of an accurate exchange-correlation 
functional the same total energies would be obtained. In the previous section, we have shown how
current-dependent terms enter the expression of the CS functional when
complex-valued orbitals are considered. Here, we evaluate the performance 
of EXX plus the CS functional in reproducing the degeneracy 
and study the effect of the current-dependent terms, in SDFT and DFT calculations. 

\begin{table}[t]
\begin{tabular}{|c|c|c|}
\hline\hline 
&\multicolumn{2}{c|}{SDFT (DFT)}\\
\cline{2-3}
~Atom~ &  ~$\Delta_{JCSLYP}$~ & ~$\Delta_{CSLYP}$   \\ [0.5ex] 
\hline 
B   &  ~0.8~(-0.3)~  & ~2.4~(1.4)~      \\ [0.5ex]
C   &  ~0.9~(-0.1)~  & ~-3.2~(-4.3)~    \\ [0.5ex]  
O   &  ~-0.6~(-1.9)~ & ~0.9~(-0.4)~     \\ [0.5ex]
F   &  ~-0.1~(-1.5)~ & ~-3.5~(-5.1)~    \\ [0.5ex]
Al  &  ~0.4~(-0.5)~  & ~1.1~(0.2)~      \\ [0.5ex]
Si  &  ~0.5~(-0.4)~  & ~-1.2~(-2.2)~	\\ [0.5ex]
S   &  ~0.1~(-1.6)~  & ~1.1~(-0.7)~ 	\\ [0.5ex]
Cl  &  ~0.7~(-1.3)~  & ~-1.1~(-3.2)~	\\ [0.5ex]
\hline 
me  &  ~0.3~(-1.0)~  & ~-0.4~(-1.8)~   \\ [0.5ex]
mae &  ~0.5~(1.0)~  & ~1.8~(2.2)~    \\ [0.5ex]
\hline\hline
\end{tabular}
\caption{\label{tab1} 
Spurios energy splittings, $\Delta = E(|M_{L}|=1)-E(M_{L}=0)$ in kcal/mol
for open-shell atoms, computed in SDFT (DFT results in parenthesis for comparison). 
Correlation energy has been added to the KLI-EXX energies 
including (JCSLYP)  and  neglecting (CSLYP) the current terms of 
Eq.~(\ref{JCSLYP}). The last row shows the mean error (me) and mean absolute  
(mae) of the spurious splittings.}
\end{table}

We consider atoms of the first and second row of the periodic table:
these are the reference cases for which a vast amount of numerical data is 
available \cite{Baerends:97,Becke:02,MaximoffErnzerhofScuseria:04,TaoPerdew:05,PittalisKurthGross:06,PittalisKurthHelbigGross:06}.
In analogy to the procedure where Hartree-Fock orbitals are used as input to the
CS formula, we have evaluated the correlation energies in a post-hoc fashion 
using Kohn-Sham (KS) orbitals. We expand the KS orbitals in   
Slater-type basis functions (QZ4P of Ref.~\cite{Velde:2001}) for the 
radial part, multiplied with spherical harmonics for the angular part.
We obtain the KS orbitals from self-consistent EXX-only
calculations employing the approximation of Krieger, Li, and Iafrate (KLI)
\cite{KriegerLiIafrate:92-2}, which has been shown to be extremely good
at least for small systems \cite{GraboKreibichKurthGross:00}.
In principle a functional should be evaluated with KS orbitals
obtained from self-consistent calculations, and this is certainly
possible for the CS functional \cite{GraboGross:95}. However, thanks to the 
variational nature of DFT, it is common experience to observe only minor
quantitative differences between post-hoc and self-consistent evaluation of total energies.
This is the reason why the functionals designed for solving
the degeneracy problem are typically evaluated in a post-hoc manner 
\cite{Becke:02,MaximoffErnzerhofScuseria:04,TaoPerdew:05}.

\vspace*{0.5cm}
\begin{figure*}[t]
\includegraphics[scale=0.5]{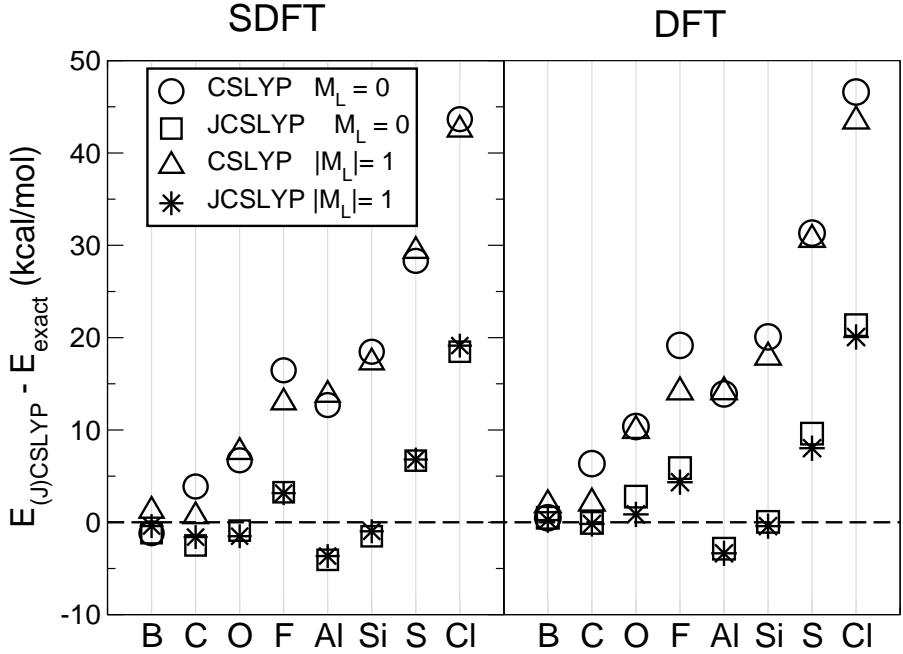}
\vspace*{0.5cm}
\caption{\label{fig1} Deviation from exact total energies for SDFT and DFT calculations employing
the CS functional, including (JCSLYP) and not including (CSLYP) the current-dependent term 
J in Eq.~(\ref{JCSLYP}).
States with different magnetic quantum numbers $M_L$ are plotted.
Exact total energies are taken from Ref. \cite{GraboGross:95} and references therein.}
\end{figure*}

Table~\ref{tab1} shows the 
spurious energy splittings (difference in the total energies) between Kohn-Sham Slater 
determinants with total magnetic quantum number $|M_{L}|=1$ and $M_{L} = 0$,
from our SDFT and DFT calculations. The deviation of these total energies from the 
exact values is plotted in Fig. (1), where again the spurious energy splittings
are visible. These results highlight the importance of including $J(\vr)$ 
in Eq.~(\ref{JCSLYP}). In particular, it is remarkable to observe that 
SDFT splittings are within 0.9 kcal/mol (with a mean error
of 0.3 kcal/mol and a mean absolute error of 0.5 kcal/mol), and 
the corresponding DFT spurious energy splittings are less 
than 1.9 kcal/mol (with mean errors of 1.0 kcal/mol). It is worthwhile to note 
that in several cases inclusion of correlation leads to current-carrying states 
($|M_{L}|=1$) with lower total energy than zero-current states ($M_{L} = 0$). 
These results are in contrast to EXX-only \cite{PittalisKurthGross:06} cases where:
(a) the zero-current states are always lowest in energy and 
(b) the spurious energy splittings are always smaller in DFT than in SDFT.

\begin{table}[t]
\begin{tabular}{|c|c|c|c|c|}
\hline\hline 
&\multicolumn{4}{c|}{JCSLYP (CSLYP)}\\
\cline{2-5}
   & ~SDFT~  & ~~DFT~~  & ~~SDFT~~  & ~~DFT~~  \\ [0.5ex] 
\hline 
~~$|M_L|$~~    & 0 &  0 &  1 &  1  \\ [0.5ex] 
\hline 
 me   &  ~~2.3~(16.1)~~     &   ~~4.7~(18.5)~~     &   ~~2.6~(15.7)~~     &   ~~3.7~(16.8)~~   \\ [0.5ex]
 mae  &  ~~4.8~(16.4)~~    &   ~~5.4~(18.5)~~     &   ~~4.7~(15.7)~~     &   ~~4.7~(16.8)~~   \\ [0.5ex]
\hline\hline
\end{tabular}
\caption{\label{tab2} 
Mean error (me) and mean absolute error 
(mae) in the total energies for 
the CS functional, including (JCSLYP) and not including (CSLYP results in parenthesis) 
the current-dependent term J in Eq.~(\ref{JCSLYP}), in kcal/mol. 
Exact total energies are taken from Ref. \cite{GraboGross:95} and references therein.}
\end{table}

Going beyond EXX by including correlation in the form of CS functional 
accurate total energies can be obtained within the OEP method \cite{GraboGross:95}.
Figure~(\ref{fig1}) and Table~\ref{tab2}
show the deviations from exact total energies for the states with different 
magnetic quantum numbers, i.e. different current-carrying states. This 
further emphasizes the importance of  proper inclusion of $J(\vr)$ in Eq.~(\ref{JCSLYP}).

\section{Conclusions}
We have shown that going
beyond the exact-exchange approximation by including correlation
energy in the form of the Colle-Salvetti functional leads to a very good description
of the degeneracy of open-shell atoms in both SDFT and DFT calculations.
Comparing DFT and SDFT results for the first and second row of the periodic table, 
on average we observe a reduction of the spurious energy splittings and better 
total energies for SDFT.
Furthermore, we have also shown how current-dependent terms enter the
expression of the Colle-Salvetti functional.
If these terms are neglected, the degeneracy is not well described and the 
total energies are also less accurate. Thus, this analysis re-confirms the 
advantage of properly including the orbital current dependent terms as an 
ingredient in correlation functionals.

\section*{Acknowledgements}
We acknowledge the
Deutsche Forschungsgemeinschaft (SPP-1145) and NoE NANOQUANTA Network (NMP4-CT-2004-50019)
for financial support.
\bibliographystyle{prstyown}

\end{document}